# Conditionally unutilized proteins and their profound effects on growth and adaptation across microbial species

Rohan Balakrishnan[1] and Jonas Cremer[2]
[1]Department of Physics, University of California at San Diego, 9500 Gilman Drive, La Jolla, CA, 92093 USA. E-mail: rbalakrishnan@ucsd.edu
[2]Department of Biology, Stanford University, 318 Campus Drive, Stanford, CA, 93105 USA. Telephone: +1 6507247178. E-mail: jbcremer@stanford.edu

**Abstract**
Protein synthesis is an important determinant of microbial growth and response that demands a high amount of metabolic and biosynthetic resources. Despite these costs, microbial species from different taxa and habitats massively synthesize proteins that are not utilized in the conditions they currently experience. Based on resource allocation models, recent studies have begun to reconcile the costs and benefits of these conditionally unutilized proteins (CUPs) in the context of varying environmental conditions. Such massive synthesis of CUPs is crucial to consider in different areas of modern microbiology, from the systematic investigation of cell physiology, via the prediction of evolution in laboratory and natural environments, to the rational design of strains in biotechnology applications.

**Introduction**
Rapid growth and quick response to changing conditions are two crucial properties that determine the fitness of microbes in their natural habitat. These two abilities may require the expression of vastly different subsets of genes. For instance, a rapid response to depleting nutrients may involve the synthesis of a diverse array of transporters to scavenge for alternative nutrient sources even before nutrients run out. In contrast, rapid growth in the currently experienced environment requires the synthesis of ribosomal and metabolic enzymes needed for protein synthesis. Given the limited capacity of cells to express genes and synthesize proteins[1–4] largely due to shared major resources like RNA polymerases and ribosomes, these different requirements for gene expression and protein synthesis can lead to strong trade-offs between growth and response [5–8]. As a result, microbes commonly harbor proteins that may not be utilized in the current condition but are helpful in other conditions. In this review, we focus on the synthesis of such *conditionally unutilized proteins* (CUP). We summarize the growing literature on the costs and benefits of conditionally unutilized proteins and discuss why their consideration is crucial to better understand cell physiology, to reveal principles of microbial evolution and adaptation, and to improve biotechnology applications.

**Proteome partitioning dictates microbial growth**
Proteins typically account for more than 50% of microbial dry mass and are resource-demanding to produce. Thus, biomass accumulation and the speed of microbial growth is strongly determined by the rate of protein synthesis, which in turn depends on the proteome composition of the cell[2–4]. In particular, efficient protein synthesis requires the abundance of ribosomal and metabolic proteins to be finetuned such that the number of active ribosomes involved in protein synthesis is high. This is illustrated well by low-dimensional allocation models[3], sketched in **Fig. 1A**, which considers how protein synthesis depends on the relative allocation of ribosomes to make different pools of proteins, including ribosomal (light blue arrow), metabolic (orange arrow), and other proteins required for growth including those involved in housekeeping functions (dark blue arrow). Remarkably, when assuming a growth-maximizing allocation of shared resources between metabolic and ribosomal protein synthesis, these models can predict major growth phenotypes observed in *E. coli* and other well-studied model organisms very well, including the ribosome content[3,9,10] and even the speed of translation with changing growth rate[11,12] (**Fig. 1B**). As such, these models emphasize that microbes balance metabolic protein and ribosome synthesis to support efficient protein synthesis. However, these considerations do not imply that cells regulate gene expression and protein synthesis to maximize growth rates in the environmental condition they experience[12]. Instead, microbes commonly synthesize conditionally unutilized proteins (CUPs) which are not required in currently experienced conditions[2,10,13–15]. Crucially, the synthesis of such proteins is costly[1–3] as it diverts shared resources like ribosomes away from the synthesis of proteins which are needed for growth. To rationalize such a seemingly wasteful synthesis of CUPs is thus a current challenge in the field[15,16]. Recent studies have started to explore the costs of CUPs synthesis in relation to their benefits by extending resource allocation models which take CUPs and their physiological functions into consideration (**Fig. 1C**, with the purple showing allocation to CUPs)[2,3,15,16]. Such modeling approaches will for example be helpful to better rationalize the observed interplay between growth behavior, the ability to respond to changing conditions such as nutrients and light, and the ability to cope with different stresses[8,10,17,18].

In addition to unutilized proteins, cells can also synthesize proteins in amounts higher than needed. Such underutilized proteins[16,19] also impose a burden on the cell and may lead to similar tradeoffs. However, the quantification of protein underutilization is challenging as it demands detailed knowledge of enzyme efficiency and abundance. In the following, we therefore focus on unutilized proteins, which by themselves already account for a large fraction of the proteome.

**The synthesis of conditionally unutilized proteins is widely prevalent across organisms**
Advancements in gene annotations and the improvements of transcriptomics, proteomics, and ribosome profiling techniques[20–24] has allowed for the systematic quantification of CUPs, revealing that they often account for a significant fraction of the proteome. For instance, the analysis of steadily growing *E. coli* indicates that CUPs account for up to half of the proteome[16,25] with their fraction changing strongly with growth conditions[22]. The analysis of proteomics data from other bacteria like *Synechocystis*[10,26] and *Cupriavidus necator*[25], the yeast *S. cerevisiae*[9,21], and the archaea *Methanococcus maripaludis*[27] support a similar picture of widespread CUP synthesis.

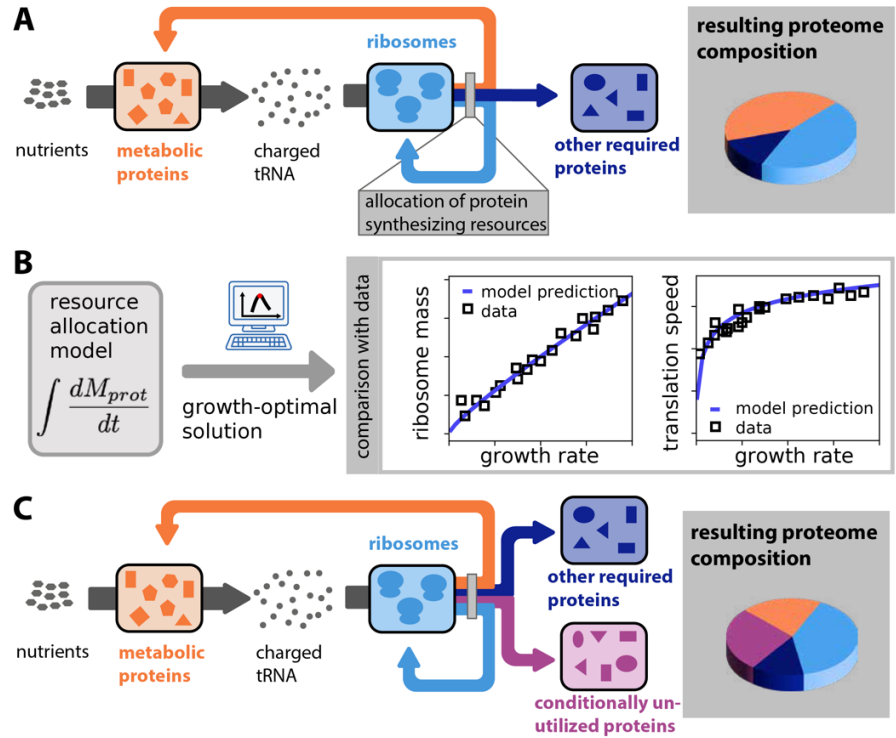

*Fig. 1: Low-dimensional allocation models connecting protein synthesis and growth. (A) Model schematics with the proteome divided into three pools of ribosomal, metabolic, and other proteins required for growth including those involved in housekeeping functions such as DNA maintenance, protein folding, etc. The rates of protein synthesis and growth are a consequence of how cells allocate ribosomes towards synthesizing the different pools with optimal growth obtained when the synthesis of ribosomal proteins is coordinated with the flux of charged tRNA metabolic proteins can provide. (B) The model can predict major growth phenotypes, including ribosome abundance (left panel) and translation speed (right panel). This predictive power emphasizes the dominant role of protein synthesis on growth as well as the cellular ability to efficiently coordinate the synthesis of ribosomal and metabolic protein pools. (C) Extension of the model to account for the synthesis of conditionally unutilized proteins which can exert their effect on growth by altering the proteome composition. This is because CUP synthesis competes for the limited pool of ribosomes thereby reducing the synthesis of other proteins.*

To explore such ubiquitous synthesis of CUPs and the resulting costs and benefits across habitats, we next discuss in more detail different organisms, with relevant findings summarized in **Table 1**.

For *E. coli*, proteins involved in amino acid biosynthesis are strongly synthesized (~3% proteome) even when growing in rich media with all required amino acids being highly abundant[27]. Furthermore, cells massively synthesize unutilized transporters and metabolic proteins when steadily growing in a poorer medium. For example, when acetate is provided as the sole carbon source, unutilized transporters specialized in the uptake of maltose, ribose, amino acids, and other carbon sources account for ~11% of the proteome[15,29] even though none of these substrates are available for uptake. Under the same condition, 14% of the proteome is also involved in motility without any apparent use in the well-shaken conditions provided during the experiments[15]. A range of perturbation experiments has further established the heavy burden and missing purpose of these proteins under the probed conditions. For example, forced synthesis of the lactose cleaving enzyme beta-galactosidase (LacZ) is of no use when lactose is not present, and severely reduces growth rates under such conditions, consistent with the high cost of producing proteins[1,3]. Conversely, the reduced synthesis of unutilized proteins increases growth- preventing the synthesis of motility proteins by deleting the motility master regulator FlhD or the major flagellin component FliC results in improved growth rates[15,29]. Such an effect is also observed in Bacillus strains lacking the expression of the motility apparatus which is unneeded yet overproduced in laboratory conditions[30]. Such strains showed increased growth rate and yield[13].

| Biological domain | Species | Typical habitat | Functional classification of CUPs |
|---|---|---|---|
| Archaea | *Methanococcus maripaludis* | Marshlands/ marine | Translation, core metabolism |
| Bacteria | *Bacillus subtilis* | Soil, ruminant gut | Motility |
| Bacteria | *Cupriavidus necator* | Soil | $CO_2$ fixation |
| Bacteria | *Escherichia coli* | Mammalian gut | Nutrient uptake, motility |
| Bacteria | *Synechocystis* PCC6803 | Freshwater | Light harvesting |
| Eukaryotes | *Saccharomyces cerevisiae* | Fruits, flowers, leaves, etc. | Respiration, translation |

*Table 1: Conditionally unutilized proteins across species and habitats. Functional classification of CUPs for a selection of microbial species. The diversity of taxa and typical habitats suggests that CUP synthesis is a ubiquitous phenotype.*

For the cyanobacteria *Synechocystis*[10], proteins of the light harvesting complex (LHC) and photosystems account for ~10% of the protein fraction, even at high light conditions and in the presence of external inorganic carbon sources in which these proteins are not needed. Accordingly, deleting antenna proteins that are part of the LHCs were found to increase the growth rate of this organism at high light conditions[31]. Conversely, under low light conditions, while a growth optimizing model would predict higher expression of light harvesting complex and photosystems genes, the cells seemed to synthesize considerable amounts of carbon catabolizing proteins instead[10].

For the lithoautotrophic soil bacterium *Cupriavidus necator*, approximately 40% of the proteome was found unlikely to be utilized in either autotrophic or heterotrophic growth conditions tested[25]. In particular, the enzymes of the Calvin-Benson-Bassham cycle, the key pathway in fixing carbon, were found to be expressed at significantly higher levels than what is predicted for optimal growth in all conditions tested. For the yeast species, *Saccharomyces cerevisiae*, cells growing in poor carbon conditions overproduce proteins involved in respiration and ribosomal proteins[9]. Under nitrogen limitation, yeast also maintains large reserves of unutilized proteins (nearly 50% proteome), especially proteins involved in translation and glucose metabolism[21]. For the archaeon *Methanococcus maripaludis*, cells maintain a roughly constant amount of ribosomal and metabolic proteins across growth conditions, despite varying demands in translational and metabolic fluxes, suggesting CUPs to account for a high fraction of the proteome [26].

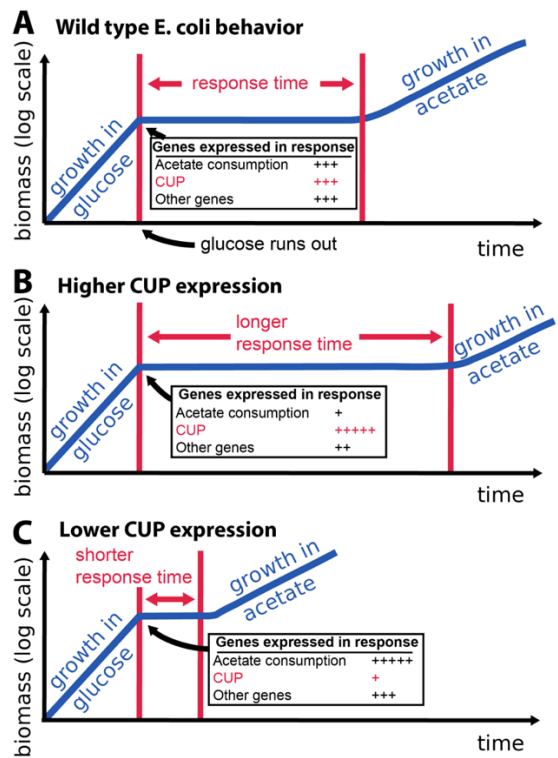

**Fig. 2: CUP synthesis delays the growth recovery of E. coli during diauxic shifts**
*When growing in batch culture with two carbon sources such as glucose and acetate, E. coli commonly utilizes glucose before acetate. (A) When glucose runs out, cells respond with the synthesis of proteins that are required to grow on acetate. But cells also synthesized proteins that are not required. It is this synthesis of CUPs that largely sets the response time before growth recovers. (B) Particularly, response times increase drastically when cells are forced to synthesize more non-required proteins during the shift, as for example realized via the artificial expression of the non-required gene lacZ. (C) In contrast, gene deletion strains that prevent the synthesis of non-needed proteins, like the motility gene fliC, show a much short response time. Gene expression measurements, summarized qualitatively in the boxes with "+" denoting the relative synthesis of different gene categories, confirm the tradeoff between the synthesis of CUPs and those required to resume growth on acetate. Adapted from Balakrishnan et al.[15].*

Besides their impact on growth in steady conditions, the synthesis of CUPs can also strongly influence the response to environmental changes. For example, following a change in nutrient availability, CUP synthesis can be very costly with the proteome taking several generations to adjust to the new environment[32]. This is well illustrated using the famous diauxic growth of *E. coli* on two carbon sources[34] wherein the long arrest of growth during the transition was shown to be largely due to an increased CUP synthesis[15] when nutrient availability changes, see **Fig. 2**.

**The benefits of CUP synthesis and adaptation to change**
The strong synthesis of CUPs across species and habitats (**Table 1**) with often adverse effects on growth begs the question of why such behavior is prevalent. The ubiquitous synthesis of CUPs suggests that it is not simply a consequence of inefficient gene regulation but rather the result of a beneficial adaptation to specific natural habitats. Particularly, when microbes encounter ever-changing conditions in their natural habitat, there is little benefit in optimizing the proteome to a particular condition. Instead, microbes synthesize proteins to best cope with changes between frequently encountered conditions, even if this involves synthesizing proteins that are not needed at a given instant [34]. While more work is needed to tightly establish the fitness benefits of CUPs in changing natural habitats, ecological considerations support this adaptation to change. For example, the massive synthesis of maltose transporters and motility genes by *E. coli* in nutrient-poor conditions mentioned above can be rationalized by the mammalian gut environment which is characterized by rapid increases in starch abundance and the flow of luminal content after meal intakes by the host[15]. Similarly, the expression of dark genes in high-light conditions by the cyanobacteria *Synechococcus* can be rationalized as a strategy to cope with varying light conditions [6]. Accordingly, repeated exposure to such specific alterations in the environment for generations can select for diversified proteomes with a large fraction of the proteome being unutilized within many conditions (**Fig. 3A**).

In line with this picture of CUP synthesis in response to environmental changes, selection in adaptive laboratory experiments under highly controlled conditions yields strains that synthesize less CUPs[16] (**Fig. 3B**). For example, when *E. coli* evolve in a well-mixed environment under rich carbon conditions[35,36], strains evolve to lose their ability to swim and grow on excluded carbon sources[37,38], functions supported by proteins which are, as mentioned above, synthesized strongly by *E. coli* even when not required. This fitness gain via reduced CUP synthesis has been observed in different controlled environments and is further supported by metabolic modeling[22]. As such it is also plausible that the simple reduction of unutilized proteins might also be the major cause of the strong diminishing return epistasis commonly reported during laboratory evolution[39–41].

**CUP synthesis in biotechnology**

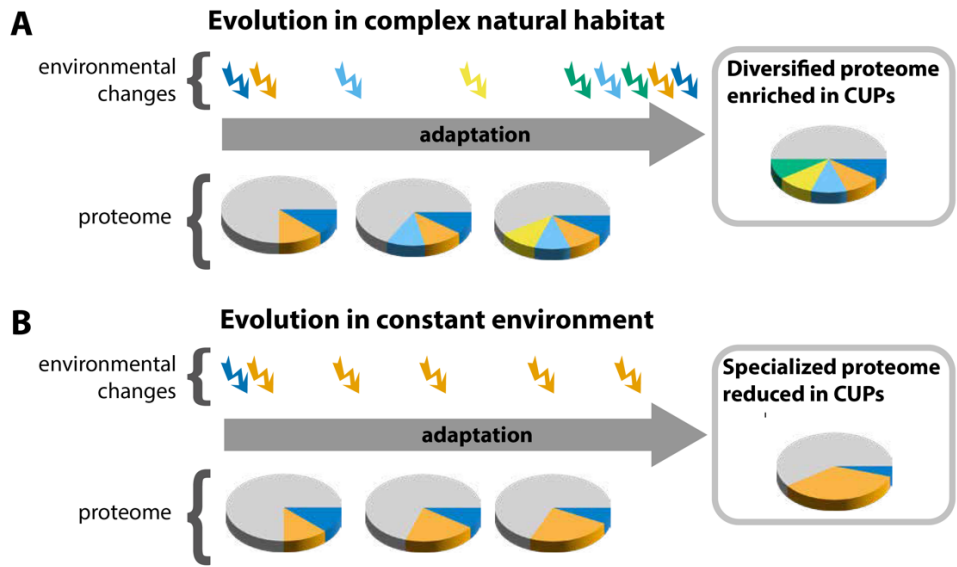

*Fig. 3. Evolutionary adaptation in complex natural habitats and simpler laboratory environments select for and against the synthesis of CUPs. Upon experiencing a specific environmental change, for example, the depletion of a carbon source (blue arrow), cells may respond by increasing specific proteins such as carbon catabolism proteins (corresponding blue sector in pie charts). (A) When evolving in complex habitats characterized by different types of changes (different colored arrows), resulting strains might thus start to synthesize a variety of proteins (pie charts, colored sectors) which help to overcome these different challenges, even in the momentary absence of any particular challenge. Consequently, a large fraction of the proteome remains unutilized in many conditions, with growth being non-optimal in those conditions. (B) In contrast, when cells evolve in constant environments (yellow arrows), for example in a lab setup with the repeated replenishment of media, the proteome becomes more optimized to cope with the specific conditions through the reduced synthesis CUPs.*

CUP synthesis and its effect on growth and biomass production are crucial to consider in biotechnology applications. Strains designed to exhibit reduced CUP synthesis could for example divert more cellular resources towards the synthesis of commercially desirable protein products, leading to increased rates and yields of production[42,43]. Over the years, strains for biotechnology applications with reduced CUP expression have been developed either through laboratory evolution runs like those mentioned above, or through rational design[42,44–46]. Adaptive laboratory evolution[47,48] utilizes natural selection in carefully designed environments to promote the appearance of beneficial traits. Strategies involve coupling the production of the protein of importance to growth, promoting the selection of high-producing evolved strains, or minimizing growth defects in engineered strains[48]. Rational design on the other hand involves manipulating cells based on prior knowledge about the various details on enzymatic functions and regulatory pathways. While we are still far from realizing such detailed characterizations [49], efforts in reducing the cellular burden by targeted gene deletion are already underway[42,50,51]. This has been realized for example in the emerging industrial workhorse *Pseudomonas putida*, particularly through deletions of its flagellar components[52].

**Conclusion**
Several studies in recent years have revealed that microbes from different taxa and habitats synthesize significant amounts of proteins not utilized in the immediate environment. This behavior may have evolved to enable rapid responses to commonly encountered variations in the natural habitat by which it potentially manifests a huge fitness advantage. But this adaptation to frequently encountered variations also leads to sub-optimal growth in other conditions. It is for example often possible to enhance growth in a given condition by reducing the synthesis of non-required proteins. Hence, studies that solely consider growth in a specific condition and rationalize cell behavior to optimally cope within this condition can only provide limited insights into microbial physiology. Rather, a successful shift of microbiology research to gain a systems-level understanding requires the tight integration of molecular, physiological, ecological, and evolutionary perspectives with the synthesis of conditionally unutilized proteins being a pivotal part. Additionally, conditionally unutilized proteins are key to consider in biotechnological applications: by manipulation the expression of non-required proteins, more cellular resources can be diverted towards desired protein products.

**Acknowledgement**
We thank Alfred Spormann and members of the Cremer, Hwa, and Spormann groups for helpful discussions on this topic. J.C. acknowledges support via a Stanford Bio-X Seeding grant and a Terman Fellowship.